\documentclass[twocolumn,aps,showpacs,preprintnumbers,nofootinbib,
superscriptaddress]{revtex4}
\usepackage{graphicx}
\usepackage{dcolumn}
\usepackage{bm}
\begin{document}

\preprint{Report at ICFA2001 WS, Stony Brook, NY, USA, 2001;
Published in physics/0111184 and presented to PRSTAB}

\title{Methods of Laser Cooling of Electron Beams in Storage Rings}

\author{E.G.Bessonov}
\affiliation{Lebedev Physical Institute RAS, Moscow, Russia}

\date{\today}

                       \begin{abstract}
Methods of enhanced laser cooling of particle beams in storage rings
and Robinson's damping criterion are discussed. The dynamics of
amplitudes of betatron oscillations and instantaneous orbits of
electrons interacting with laser beams being displaced in the radial
direction is investigated.  \end{abstract}


\maketitle

                     \section{Introduction}

Different cooling methods were suggested to decrease the emittances and
to compress the phase-space density of charged particle beams in
storage rings. All of them, except the stochastic one, are based on a
friction of particles in external electromagnetic fields or in media.
Below we will consider laser methods of enhanced cooling of electron
beams based on friction. The main results will be valid for cooling of
other particles.

       \section{Methods of cooling of particle beams in storage rings}

In the ordinary three-dimensional radiative method of cooling of
electron beams a laser beam overlaps an electron beam, its
transverse position is motionless, all electrons interact with the
laser beam independent of their energy and amplitude of betatron
oscillations. A friction originating in the process of emission
(scattering) of laser photons by electrons leads to a damping of
amplitudes of both betatron and phase oscillations of electrons. The
damping is because of the friction force is parallel to the electron
velocity, and therefore the momentum losses include both the transverse
and longitudinal ones. The longitudinal momentum losses are compensated
by a radio frequency accelerating system of the storage ring. Meanwhile
the longitudinal momentum of the electron tends to a certain
equilibrium. The transverse vertical and radial momenta disappear
irreversibly. The difference in rates of momentum loss of electrons
having maximum and minimum energies in the beam is small, and that is
why the cooling time of the electron beam is high.

When a cooling is produced in a dispersion-free straight section and the
laser beam intensity is constant inside the area of the laser beam
occupied by a being cooled electron beam, then the damping times of the
horizontal vertical and phase oscillations are:

        \begin{equation}   
        \tau _x = \tau _y = 2\tau _{\epsilon} =
        {2\varepsilon \over \overline P},
        \end{equation}
where $\overline P$ is the average power of the radiation scattered by
the electron of the energy $\varepsilon$.

The coupling arising in non-zero dispersion straight sections of the
storage rings leads to a redistribution of the longitudinal and radial
damping times when the radial gradient of the laser beam intensity is
introduced.

The physics of three-dimensional radiative cooling both ion and
electron beams is not differ from synchrotron radiation damping. We
deal with the non-conservative system. Friction causes the appearance of
a reaction force on the emitting particle which must be taken into
account to describe particle dynamics. The Liouville's theorem does not
valid for such systems. At the same time in the case of non-selective
interaction the Robinson's damping criterion is valid: the sum of
damping decrements (inverse damping times) is a constant ($\tau _y
^{-1} + \tau _s ^{-1} = 3 \overline P/2\varepsilon$). In a particular
choice of lattice and laser or material target, damping rates can be
shifted between different degrees of freedom. However the maximum
decrements are limited by the condition $\tau _{y,s} \geq 0$ or by the
value $\tau _{y,s} \geq 2\varepsilon /3 \overline P$ \cite {wiedemann,
ICFA01}.

The method of enhanced one-dimensional laser cooling of ion beams in
the longitudinal plane is based on the resonance Rayleigh scattering of
"monochromatic" laser photons by not fully stripped ion beams or by
complicated nuclei.  In this method the laser beam overlaps the ion
beam and has a chirp of frequency. Ions interact with the laser beam at
resonance energy, decrease their energy in the process of the laser
frequency scanning until all of them reach the minimum energy of ions
in the beam. At this frequency the laser beam is switched off. The
higher energy of ions the longer the time of interaction of ions with
the laser beam. Ions of minimum energy do not interact with the laser
beam at all. In such a way, in this method, the selective interaction
is realized. The damping time of the ion beam in the longitudinal plane
is determined by the dispersion of its energy spread $\sigma
_{\varepsilon}$

        \begin{equation}   
        \tau _{\epsilon} = {2 \sigma _{\varepsilon} \over
        \overline P}. \end{equation}

According to (1), the damping times of betatron and phase oscillations
in the three-dimensional radiative method of cooling are equal to the
time interval necessary for the electron energy loss, which is about the
two-fold and four-fold initial energy of the electron, accordingly. The
damping time (2) is $\varepsilon / \sigma _{\varepsilon} \sim 10 ^3$
times less than (1). It means, that the selectivity of interaction
leads to the violation of the Robinson's damping criterion and open the
possibility of enhanced cooling.

A one-dimensional method of ion cooling has been realized by now \cite
{channel} - \cite {hangst1}. A three-dimensional radiative method of
laser cooling of ion beams by broadband laser beams was suggested and
developed in \cite {idea} - \cite {pac} by the analogy with the
synchrotron radiation damping of amplitudes of betatron and phase
oscillations of particles in storage rings\footnote{The difference in
cooling of electron and ion beams is in the dependence of the average
power of scattered radiation on the relative energy ($\overline P _e
\sim \gamma ^2$, $\overline P _{ion} \sim \gamma$) and in the finite
decay time of ions excited in the process of the Rayleigh scattering.
The latter may be neglected when length of the ion decay is much less
than that of the period of ion betatron oscillations \cite {idea},
\cite {prl}.}.  The electron version of the three-dimensional radiative
cooling was developed by Zh.Huang and R.D.Ruth \cite {zhirong}. They
paid attention to the possibility to store laser wavepackets of
picosecond duration and high intensity, $I \simeq 10 ^{17}$ W/cm$^{2}$
(magnetic field strength, $B \sim 2 \cdot 10 ^7$ Gs) in the optical
high-finesse resonator to interact repetitively with a circulating
electron beam in a storage ring of the energy $10\div10 ^2$ MeV for the
rapid cooling of the beam, counterbalancing of the intrabeam
scattering, and x-ray generation. Below the enhanced methods of
radiative laser cooling of electron beams in storage rings are
discussed \cite {bkw}, \cite {bes2}.

\section {Enhanced laser cooling of electron beams}

Below the enhanced cooling of particle beams both in the longitudinal
and transverse planes will be considered. A universal kind of selective
interaction of particles with moving in the radial direction laser
beams or media targets is suggested. Selective interaction means that
the laser beam or media target at some moment interacts with one part
of the being cooled beam and does not interact with the rest. At the
other moments the interaction is switched on for another parts of the
being cooled beam. Robinson's damping criterion does not work in this
case.

          \subsection {Cooling methods based on selective interaction
          of electron and laser beams and on dispersion coupling of the
          transverse and longitudinal\\ motion of electrons in storage
          rings}

For the sake of simplicity we will neglect the emission of the
synchrotron radiation in the bending magnets of the storage rings,
supposing that the RF system of a storage ring is switched off and the
laser beams are homogeneous and have sharp edges in the radial
directions.  We suppose that the dispersion of the energy loss of
electrons in the laser beam is small and the jump of their
instantaneous orbits caused by the energy loss is less than the
amplitude of betatron oscillations.

In a smooth approximation, the movement of an electron relative to its
instantaneous orbit is described by the equation

       \begin{equation}             
       x _{\beta} = A\cos(\Omega t+\varphi).
       \end{equation}
where $x _{\beta} = x - x_{\eta}$ is the electron deviation from the
instantaneous orbit $x_{\eta}$; $x$, its radial coordinate; $A$ and
$\Omega$, the amplitude and the frequency of betatron oscillations.

If the coordinate $x _{\beta\,0}$ and transverse radial velocity of the
electron $\dot x _{\beta\,0} = - A \Omega \sin (\Omega t _0 + \varphi
)$ correspond to the moment $t _0$ of change of the electron energy in
a laser beam then the amplitude of betatron oscillations of the
electron before an interaction is $A _0 = \sqrt {x _{\beta\,0} ^2 +
\dot x _{\beta \,0} ^2 /\Omega ^2}$. After the interaction, the
position of the electron instantaneous orbit will be changed by a value
$\delta x_{\eta}$, the deviation of the electron relative to the new
orbit will be $x _{\beta \,0} - \delta x _{\eta}$, and the change of
its transverse velocity can be neglected\footnote{It means that we
neglect a week ordinary damping of amplitudes of betatron oscillations
determined by a loss of the transverse electron momentum.}. The new
amplitude of the electron betatron oscillations will be $A_1 =\sqrt {(x
_{\beta \,0} - \delta x _{\eta})^2 + \dot x ^2 _{\beta \,0} /\Omega
^2}$ and the change of the square of the amplitude

       \begin{equation}
       \delta (A)^2 = A_1^2 - A_0^2 = - 2x _{\beta \,0}\delta x _{\eta}
       + (\delta x _{\eta}) ^2.
       \end{equation}          

When $|\delta x _{\eta}| \ll |x _{\beta\,0}| < A _0$ then in the first
approximation the value $\delta A = - (x _{\beta \,0} /A) \delta x
_{\eta}$. From this it follows that to produce the enhanced cooling of
an electron beam in the transverse plane we must create such conditions
when electrons interact with a laser beam under deviations from the
instantaneous orbit $x _{\beta 0}$ of one sign ($x _{\beta \,0} < 0$,
when the dispersion function is positive $\partial x _{\eta}/ \partial
\varepsilon > 0$ or in the opposite case $x _{\beta \,0} > 0$,
$\partial x _{\eta}/ \partial \varepsilon < 0$). \footnote {Friction
in non-conservative system can lead to cooling in one plane and to
heating in the other one.}. In this case the value $\delta A $ has
one sign and the rate of change of amplitudes of betatron oscillations
of electrons $\partial A /\partial t$ is maximum.  A selective
interaction of electrons with the laser beam is necessary to realize
this case.

In Fig.1, two schemes of a selective interaction of electron and laser
beams are shown for cooling of electron beams in the transverse and
longitudinal planes. For the transverse cooling, the laser beam $T _1$
is used. At the initial moment it overlaps a small external part of the
electron beam in the radial direction in the straight section of the
storage ring with non zero dispersion function. First electrons with
largest initial amplitudes of betatron oscillations interact with the
laser beam. Immediately after the interaction and loss of the energy
the position and direction of momentum of an electron remain the same,
but the instantaneous orbit is displaced inward in the direction of the
laser beam. The radial coordinate of the instantaneous orbit and the
amplitude of betatron oscillations are decreased to the same value
owing to the dispersion coupling. After every interaction the position
of the instantaneous orbit approaches the laser beam more and more, and
the amplitude of betatron oscillations is coming smaller.  It will
reach some small value when the instantaneous orbit reaches the edge of
the laser beam.  When the depth of dipping of the instantaneous orbit
of the electron in the laser beam becomes greater than the amplitude of
its betatron oscillations, the orbit will continue its movement in the
laser beam with constant velocity. The amplitude of betatron
oscillations will not be changed.

\vskip 0mm
\begin{figure}[hbt]
\includegraphics{
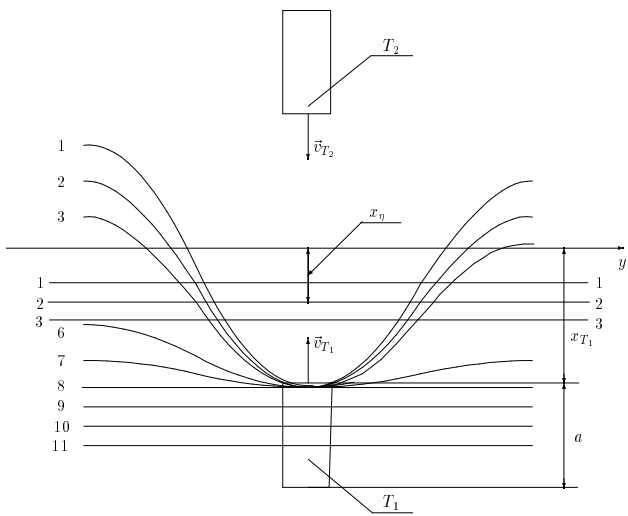}
\caption{\label{fig:epsart} The scheme of the enhanced electron
cooling. \\The axis $"y"$ is the equilibrium orbit of the storage ring;
$T_1$ and $T_2$ the laser beams. The transverse positions of laser beams
are displaced with the velocity $\vec v _{T _{1,2}}$ relative to the
equilibrium orbit, 1-1, 2-2, ...  the location of the instantaneous
electron orbit, and 1,2,3, ... the electron trajectories after 1,2,3,
...  events of the energy loss.} \end{figure}

\vskip -1mm

The degree of overlapping is changed by moving uniformly the laser beam
position from inside in the direction of the being cooled electron beam
with some velocity $v _{T _1}$\footnote {Instantaneous orbits can be
moved in the direction of the laser beam, instead of moving of a laser
beam.  A kick, decreasing of the value of the magnetic field in bending
magnets of the storage ring, a phase displacement or eddy electric
fields can be used for this purpose.}.  When the laser beam reaches the
instantaneous orbit corresponding to electrons of maximum energies then
the laser beam must be switched off and returned to a previous
position. All electrons of the beam will have small amplitudes of
betatron oscillations and increased energy spread. Electrons with high
amplitudes of betatron oscillations will start to interact with a laser
beam first, their duration of interaction and absolute decrease of
amplitudes of betatron oscillations will be higher.

To realize the enhanced cooling of an electron beam in the longitudinal
plane we can use a laser beam $T _2$ located in the straight section of
a storage ring with non zero dispersion function (see Fig.1). The
radial laser beam position is moving uniformly from outside in the
direction of the being cooled electron beam with a velocity $v _{T _2}$
higher than maximum velocity of the electron instantaneous orbit
deepened in the laser beam. At the initial moment, the laser beam
overlaps only a small part of the electron beam. The degree of
overlapping is changed in such a way that electrons of maximum energy,
first and then electrons of lesser energy, come into interaction. When
the laser beam reaches the orbit of electrons of minimum energy then it
must be switched off and returned to the previous position. In this
case, the rate of the energy loss of electrons in the beam will not be
increased, but the difference in duration of interaction and hence in
the energy losses of electrons having maximum and minimum energies will
be increased essentially. As a result all electrons will be gathered at
the minimum energy in a short time.

      \subsection {Interaction of electron beams with transversely
      moving laser beams}

In the methods of enhanced laser cooling of electron beams the internal
and external laser beam positions are displaced in the transverse
directions (see Fig.1). Below the evolution of amplitudes of betatron
oscillations and positions of instantaneous orbits in the process of
the energy loss of electrons in laser beams will be analyzed.

The velocity of an electron instantaneous orbit $\dot x_{\eta}$ depends
on the distance $x _{T _{1,2}} - x _{\eta}$ between the edge of the
laser beam and the instantaneous orbit, and on the amplitude of
betatron oscillations. When the orbit enters the laser beam at the
depth higher than the amplitude of betatron oscillations then electrons
interact with the laser beam every turn and theirs velocity reaches the
maximum value $\dot x_{\eta \, in}$ which is given by the intensity and
the length of the interaction region of the electron and laser beams.
In the general case, the velocity $\dot x_{\eta}$ can be presented in
the form $\dot x_{\eta} =  W\cdot \dot x_{\eta \, in}$, where $W$ is
the probability of an electron crossing the laser beam. $W$ is the
ratio to a period of a part of the period of betatron oscillations of
the electron determined by the condition $| x _{T _{1 2}} - x_{\eta}|
\leq |x _{0}| \leq A$ when the deviation of the electron from the
instantaneous orbit at azimuths of the laser beam is directed to the
laser beam and is greater than the distance between the orbit and the
laser beam. When the length of the laser beam is much less than the
length of the period of electron betatron oscillations, the
probability can be presented in the form $W = \varphi _{1, 2}/\pi$,
where $\varphi _{1} = \pi - \arccos \xi _{1}$, $\varphi _{2} = \arccos
\xi _{2}$, $\xi _{1, 2} = (x _{T_{1, 2}} - x _{\eta}) /A$, indices 1,2
correspond to laser beams.

The behavior of the amplitudes of betatron oscillations of electrons,
according to (4), is determined by the equation $\partial A/\partial x
_{\eta} = -<x _{\beta\,0}>/A$, where $<x _{\beta\,0}>$ is the electron
deviation from the instantaneous orbit averaged through the range of
phases $2 \varphi _{1, 2}$ of betatron oscillations where electrons
cross the laser beam. The value $<x _{\beta\,0}> = \pm A sinc \,
\varphi _{1 ,2}$, where $sinc \varphi _{1, 2} = sin \varphi _{1,
2}/\varphi _{1, 2}$, signs $+$ and $-$ are related to the first and
second laser beams. Thus the cooling processes are determined by the
system of equations

       \begin{equation}               
       {\partial A\over \partial x_{\eta}} = \pm sinc \varphi
       _{1,2}, \hskip 5mm {\partial x _{\eta}\over \partial t} = {\dot
       x_{\eta \, in} \over \pi}\varphi _{1,2}.
       \end{equation}

From equations (5) and the expression $\partial A/\partial x _{\eta} =
[\partial A/\partial t]/ [\partial x _{\eta}/\partial t]$ it follows:

       \begin{equation}                                
       {\partial A\over \partial t} = {\dot x _{\eta \,in}\over
       \pi}\sin\varphi _{1,2} = {\dot x _{\eta \,in}\over
       \pi}\sqrt {1 - \xi _{1,\,2}^2}.  \end{equation}

Let the initial instantaneous electron orbits be distributed in a
region $\pm \sigma _{x, \varepsilon, 0}$ relative to the location of
the middle instantaneous orbit $\overline {x _{\eta}}$, and the initial
amplitudes of electron radial betatron oscillations $A _0$ be
distributed in a region $\sigma _{x,b,0}$ relative to their
instantaneous orbits, where $\sigma _{x, \varepsilon, 0}$ and $\sigma
_{x,b,0}$ are dispersions. The dispersion $\sigma _{x, \varepsilon, 0}$
is determined by the initial energy spread $\sigma _{\varepsilon, 0}$.

Suppose that the initial spread of amplitudes of betatron oscillations
of electrons $\sigma _{x,b,0}$ is identical for all instantaneous
orbits of the beam. The velocities of the instantaneous orbits in a
laser beam $\dot x _{\eta \, in} < 0$, the transverse velocities of the
first laser beams $v _{T _{1}} > 0$, and $v _{T _{2}} < 0$. Below we
will use the relative radial velocities of the laser beam displacement
$k _{1,2} = v _{T _{1,2}}/\dot x _{\eta \, in}$, where $v _{T _{1,2}} =
dx _{T _{1,2}}/dt$.  In our case $\dot x_{\eta \, in} <0$, $k _1 < 0$,
$k _2 > 0$, $|k _{1,2}| < 1$.

From the definition of $\xi _{1, 2}$ we have a relation $x _{\eta} = x
_{T _{1,2}} - \xi _{1, 2} A(\xi _{1, 2})$. The time derivative is
$\partial x _{\eta}/\partial t = v _{T _{1,2}} - [A + \xi _{1,2}
(\partial A/ \partial \xi _{1,2})] \partial \xi _{1,2} /\partial t$.
Equating this value to the second term in (5) we will receive the time
derivative

        \begin{equation}                               
        {\partial \xi _{1,2}\over \partial t} = {\dot x _{\eta \,in}
        \over \pi} {\pi k _{1,2} - \varphi _{1, 2} \over A(\xi
        _{1,2}) + \xi _{1,2} (\partial A / \partial \xi
        _{1,2})}.  \end{equation}

Using this equation we can transform the first value in (5) to the form
$\pm sinc \varphi _{1,2} (\xi _{1,2}) = ({\partial A/ \partial \xi _{1,
2}}) ({\partial \xi _{1,2}/ \partial t})/$ $ ({\partial x _{\eta}/
\partial t}) = $ $(\pi k _{1,2} - \varphi _{1,2}) {(\partial A
/\partial \xi _{1,2})}/ [A + \xi _{1,2}(\partial A/$ $\partial \xi
_{1,2})]\cdot $ $\varphi _{1,2}$ which can be transformed to $\partial
\ln A / \partial \xi _{1,2} = \pm \sin \varphi _{1,2} / \pi k _{1,2} -
(\varphi _{1,2} \pm \xi _{1,2} \sin \varphi _{1,2}).$ The solution of
this equation is

       \begin{equation}                         
       A = A _0 \exp \int _{\xi _{1,2,0}} ^   {\xi _{1,2}}
       {\pm \sin \varphi _{1,2} d\xi _{1,2}
       \over \pi k _{1,2} - (\varphi _{1,2}\pm \xi _{1,2}
       \sin \varphi _{1,2})},
       \end{equation}
where the index $0$ correspond to the initial time. Substituting the
values $A$ and $\partial A/ \partial \xi _{1,2}$ determined by (8) in
(7) we find the relation between time of observation and parameter
$\xi _{1,2}$

       \begin{equation}                                
       t - t _0 = {\pi A _0 \over |\dot x _{\eta \,in}|} \psi
       (k _{1,2}, \xi _{1,2}),   \end{equation}
where $\psi (k _{1,2}, \xi _{1,2}) = - \int ^{\xi _{1,2 }}
_{\xi _{1,2,  0}} A(\xi _{1,2 })/ \{A_0 [ \pi k _{1, 2} - $
$ (\varphi _{1,2} \pm \xi _{1,2 }\sin \Delta \varphi
_{1,2})]\}d\xi _{1,2 }. $

The equations (9) determine the time dependence of the functions
$\xi _{1,2}(t  - t _0)$. The dependence of the amplitudes
$A[\xi _{1,2} (t - t _0)]$ is determined by the equation (8)
through the functions $\xi _{1,2}(t - t_0)$ in a parametric form. The
dependence of the position of the instantaneous orbit follows from
the definition of $\xi _{1,2}$

       $$x_{\eta}(t - t _0) = x_{T _{1, 2}0} + v _{T_{1,2}}(t  -
       t_0) - $$
\vskip -7mm
       \begin{equation}                            
       A[(\xi _{1, 2} (t  - t_0)] \cdot \xi (t  - t _0).
       \end{equation}

The function $\psi (k _2,\xi _{2})$ for the case $k _2 > 0$ according
to (9) can be presented in the form

       $$\psi (k _2, \xi _{2}) = \int _{\xi _{2}} ^{1}{dx\exp
       \int ^{1} _{x}} $$

       \begin{equation}             
       {{\sqrt{1 - t^2} / (\pi k _{2} - \arccos t +
       t\sqrt {1 - t ^2})} \over \pi k _{2} - \arccos x + x \sqrt
       {1 - x ^2}}d\,t.
       \end{equation}

The instantaneous orbits of electrons having initial amplitudes of
betatron oscillations $A _0$ will be deepened into the laser beam to
the depth greater than their final amplitudes of betatron oscillations
$A _f$ at a moment $t _f$.  According to (9), $t _f = t _0 + \pi A _0
\psi (k _2, \xi _{2,f})/|x _{\eta \, in}|$, where $\xi _{2,f} = \xi (t
_f) = 1$. During the interval $t _f - t _0$ the laser beam $T _2$ will
pass a way $l _f = |v _{T _2}|(t _f - t _0) = \pi k_2 \psi (k _2, \xi
_{2,f}) A _0$.  The dependence $\psi (k _2, \xi _{2, f})$ determined by
(11) is presented in Table 1.

\vskip 2mm
\begin{figure}[hbt]
Table 1
\hskip 20 mm \vskip 2mm
\begin{tabular}{|l|l|l|l|l|l|l|l|l|l|l|l|l|l|}
\hline
$k _2$ & 1.0 & 1.02 & 1.03 & 1.05 & 1.1&1.2 &1.3&1.4&1.5&1.7&2.0 \\
\hline
$\psi $
&$\infty$&13.8&9.90&6.52&3.71&2.10&1.51&1.18&.98& .735&.538\\
\hline \end{tabular}

\end{figure}

\vskip 2mm

Numerical calculations of the dependence $\psi (k _2, \xi _{2})$ on
$\xi _{2}$ for the cases $k _2 = 1.0$, $k _2 = 1.1$ and $k _2 = 1.5$
are presented in Tables 2, 3, and 4, respectively. It can
be presented in the next approximate form

       \begin{equation}         
       \psi (k _2, \xi _{2}) \simeq C _3(k _2)\psi ({1 - \xi _{2}
       \over k _2 + \xi _{2}}),
       \end{equation}
where $C _3(k _2) \simeq 0.492 - 0.680 (k _2 -1) + 0.484(k _2 - 1)^2 +
...$, $\psi [(1 - \xi _{2}) / (k _2 + \xi _{2})]|_{k _2 = 1}
\simeq (1 - \xi _{2}) / (1 + \xi _{2})$.

\vskip4mm
\begin{figure}[hbt]
Table 2 \hskip 20 mm ($k_2 = 1.0$)
\vskip 2mm
\begin{tabular}{|l|l|l|l|l|l|l|l|l|l|l|}
\hline
$\xi _{2}$ & 1.0 & 0.5 & 0.2& 0&-0.2&-0.5&-0.8&-0.9& -1.0 \\
\hline
$\psi $ & $0$ &.182&.341&.492&.716& 1.393&4.388 &10.187&-$ \infty $\\
\hline \end{tabular}

\vskip 10mm
Table 3 \hskip 20 mm ($k_2 = 1.1$)
\vskip 2mm
\begin{tabular}{|l|l|l|l|l|l|l|l|l|l|l|}
\hline
$\xi _{2}$ & 1.0 & 0.5 & 0.2&0 &-0.2&-0.5& -0.8 & -0.9& -1.0  \\
\hline
$\psi $&$0$&.163&.300&.423&.595&1.033& 2.076&2.759
& 3.710 \\
\hline \end{tabular}

\vskip 10mm
Table 4  \hskip 20 mm ($k_2 = 1.5$)
\vskip 2mm
\begin{tabular}{|l|l|l|l|l|l|l|l|l|l|l|l|l|l|}
\hline
$\xi _{2}$ & 1.0 & 0.5 & 0.2&0 & -0.2& -0.4&-0.6&-0.8&-1.0  \\
\hline
$\psi )$&$0$&0.116&0.202&0.273&0.359& 0.466 &0.602 &0.772 &0.980 \\
\hline
\end{tabular}
\end{figure}
\vskip 10mm

\subsection {The enhanced transverse laser cooling of electron beams}

In the method of the enhanced transverse laser cooling of electron
beams a laser beam $T_1$ is located in the region ($x _{T _1}$, $x_{T
_1} - a$), where $a$ is the laser beam width (Fig.\,1). The degree of
the transverse cooling of the electron beam is determined by (8). The
final amplitude in this case can be presented in the form

       \begin{equation}           
       A _f = A _0 exp {\int _{\xi _{1,0}} ^{\xi _{1,f}}
       { \sqrt{1 - \xi _1^2} d\xi _1 \over \pi k _{1} - \pi +
       \arccos \xi _1 - \xi _1 \sqrt {1 - \xi _1 ^2}}}.
       \end{equation}

The numerical calculations of the dependence of the ratio $A _f /A _0$
on the relative radial velocity $k _1$ of the laser beam displacement
are presented in Fig.2 and Table 5 for the case $\xi _{1,0} = - 1$,
$\xi _{1,f} = 1$. This dependence can be presented by the approximate
expression

       \begin{equation}          
       A _{f} \simeq A _0 \sqrt{|k_1|\over |k _1| +1}.
        \end{equation}

\begin{figure}[hbt]
\includegraphics{
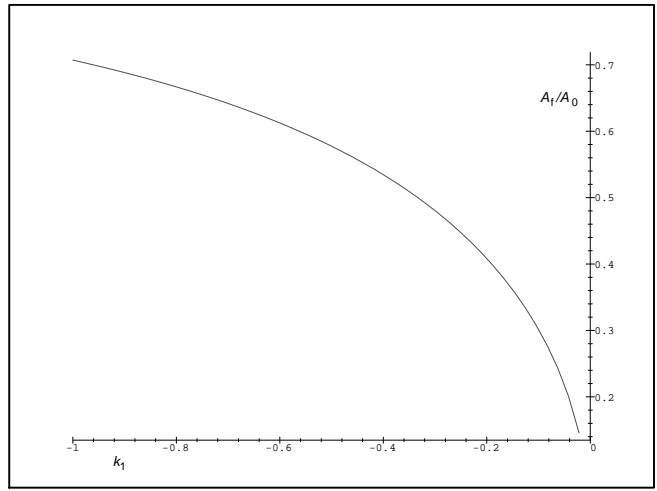}
\caption{\small \it The dependence of the ratio $A _{f}/A_0$ on $k _1$.}
\vskip 4mm

Table 5
\vskip 5mm
\begin{tabular}{|l|l|l|l|l|l|l|}
\hline
$|k_1|$ &0&0.2&0.4&0.6&0.8&1.0\\
\hline
$A_{f}/A_0$  &0&0.408&0.535&0.612&0.667&0.707\\
\hline
\end{tabular}
\end{figure}

The time of the laser beam cooling and the final total radial dimension
of the beam are equal to

         \begin{equation}                
         \tau _{x,1} \simeq {\sigma _{x, 0} \over v _{T _1}},
         \hskip 5mm \sigma _{x,f}|_{|k _1| \ll 1} \simeq {\sigma _{x,0}
         \over |k _1|} + \sigma _{x, \varepsilon, 0},
         \end{equation}
where $\sigma _{x,0} = \sigma _{x,b,0} + \sigma _{x, \varepsilon,0}$ is
the total initial radial dimension of the electron beam. For the time
$\tau _{x,1}$ the instantaneous orbits of electrons of a beam having
minimum energy and maximum amplitudes of betatron oscillations at $|k
_1| \ll 1$ pass the distance $\sim |\dot x _{\eta\, in}| \tau _{x,1}$.

According to (14) and (15) the enhanced transverse laser cooling can
lead to an appreciable degree of cooling of electron beams in the
transverse plane and a much greater degree of heating in the
longitudinal one. From this it follows that the four- and
six-dimensional emittances of the beam will be increased. Straight
sections with low-beta and high dispersion functions have to use. In
this case the radial beam dimension (15) will be lesser and lesser
events of photon emission are required to cool the beam in the
transverse direction, as the change of amplitudes of betatron
oscillations of electrons is of the order of the change of positions of
their instantaneous orbits.  Meanwhile, the spread of amplitudes of
betatron oscillations is small and the step between positions of
instantaneous orbits is high.

Similar method of interaction of external and internal targets with
proton beams was described in 1956 by O'Neil \cite{oneil,shoch}.
However, O'Neil considered the question of damping of betatron
oscillations of proton beams in the transverse direction by means of
motionless solid wedge-shaped material targets. The targets in that
case could not provide any enhanced cooling\footnote{An internal target
could be rotated out of the medium plane only to prevent the proton
beam losses when the positions of instantaneous orbits reached the edge
of the storage ring.}. They could be used for injection and capture of
only one portion of protons.  For the multi-cycle injection and storage
of protons O'Neil suggested an ordinary three-dimensional ionization
cooling based on a thin hydrogen jet target located in a working region
of the storage ring.

\subsection {The enhanced longitudinal laser cooling of electron beams}

In the method of the enhanced longitudinal laser cooling of electron
beams a laser beam $T _2$ is located in the region ($x _{T _2}$, $x_{T
_2} + a$) (see Fig.\,1). Its radial position is displaced uniformly
with the velocity $v _{T _2} < 0$, $|v _{T _2}| > |\dot x _{\eta \, in}|$
from outside of the working region of the storage ring in the direction
of a being cooled electron beam. The instantaneous orbits of electrons
will go in the same direction with a velocity $|\dot x _{\eta}| \leq
|\dot x _{\eta \, in}|$ beginning from the moment of their first
interaction with the laser beam. When the laser beam reaches the
instantaneous orbit of electrons having minimum initial energies it
must be removed to the initial position.

The law of change of the amplitudes of electron betatron oscillations
is determined by (8), which can be presented in the form

       \begin{equation}                                 
       A  = A _0 exp {\int _{\xi _{2, 0}} ^{\xi _{2}}{- \sqrt{1
       - \xi _2^2} d\xi _2 \over \pi k _{2} - \arccos \xi _2+ \xi
       _2\sqrt {1 - \xi _2^2}}}.  \end{equation}

The dependence of the ratio of a final amplitude of electron betatron
oscillations $A _f = A(\xi _2 = 1)$ to the initial one on the relative
velocity $k _2$ of the second laser beam is presented in Table 6 and
Fig.3. This ratio can be presented by the next approximate expression

       \begin{equation}                            
       A _{f} \simeq  A _0\sqrt {k_2\over k_2 - 1}.
        \end{equation}

\vskip 6mm
Table 6
\vskip 5mm
\begin{tabular}{|l|l|l|l|l|l|l|}
\hline
$k_2$ & 1.0001 & 1.001& 1.01 & 1.1& 1.5 & 2.0 \\
\hline
$A _{f}/A _0$ &$100.005$&31.64&10.04 &3.32& 1.73&1.414 \\
\hline
\end{tabular}

\vskip 5mm
The evolution of instantaneous orbits of electrons interacting with the
laser beam depends on the initial amplitudes of betatron oscillations
of these electrons. First of all the laser beam $T _2$ interacts with
electrons having the largest initial amplitudes of betatron
oscillations $A _0 = \sigma _{x,b,0}$ and the highest energies. The
instantaneous orbit of these electrons, according to (8) - (10), is
changed by the law $ x _{\eta _1} = x _{T _2, 0} + v _{T _2}(t - t _{0
_1}) - \xi _{2} \sigma _{x,b}(\xi _{2})$ up to the time $t = t _f$,
where $t _{0 _1}$ is the initial time of interaction of electrons with
the laser beam. At the same time instantaneous orbits $x _{\eta _2}$ of
electrons having the same maximum energy but zero amplitudes of
betatron oscillations are at rest up to the moment $t _{0 _2} = t _{0
_1} + \sigma _{x,b,0}/|v _{T_2}|$. The orbit $x _{\eta _1}$ is
displaced relative to the orbit $x _{\eta _2}$ by the distance $\Delta
x _{\eta _{1-2}} = (x _{\eta _1} - x _{\eta _2})$. At the moment $t _{0
_2}$, when $x _{\eta _2} = x _{T _2}$, this distance reaches the minimum

       \begin{equation}
       \Delta x _{\eta _{1-2} m}(t _{0 _2}) =
       - \xi _2(t _{0 _2})\cdot \sigma _{x,b}(t _{0 _2}) < 0,
       \end{equation} 
where the parameter $\xi _2(t _{0 _2})$, according to (9) and the
condition $|v _{T _2}|(t _{0 _2} - t _{0 _1}) = \sigma _{x,b,0}$, will
be determined by the equation $\psi [k _2, \xi _{2}(t _{0 _2})] = 1/\pi
k_2$. The value $\psi [k _2, \xi _{2}(t _{0 _2})]| _{k_2 \simeq 1}
\simeq 1/\pi$, $\xi _{2} (k _2, t _{0 _2})|_{k _2 \simeq 1}$ $ \simeq
0.22$ (see Tables 2-4), $\sigma _{x,b}(t _{0 _2}) = 1.26 \sigma
_{x,b,0}$ and the distance $|\Delta x _{\eta _{1-2}}(t _{0 _2})| \simeq
0.28 \sigma _{x,b,0}$. This distance is decreased with increasing $k
_2$.

\vskip 30mm
\begin{figure}[hbt]
\includegraphics{
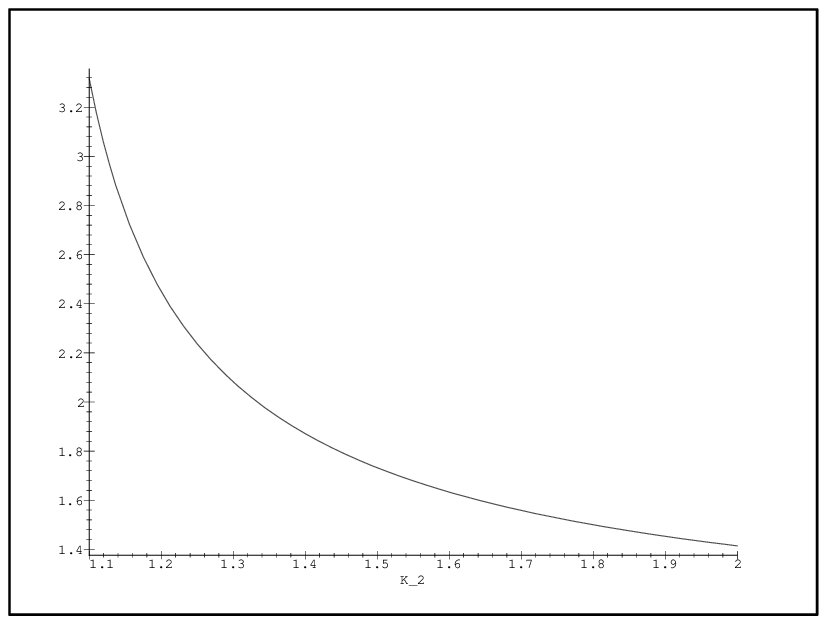}
\caption{\small \it The dependence of the ratio $A _{f}/A_0$ on $k
_2$.}
\vskip -60mm
\hskip -40mm
{\small $A _{f}/A_0$}
\end{figure}
\vskip 55mm

The instantaneous orbit of particles $x _{\eta _2}$ inside the interval
$t _{0 _2} < t \leq t _f$ is changed by the law $x _{\eta _2} = x _{T
_2, 0}$ $ - \sigma _{x,b,0} + \dot x _{\eta,\,in}(t - t _{0 _1} +
\sigma _{x,b,0}/v _{T_2})$ and the distance

           $$\Delta x _{\eta _{1- 2}} = {(k _2 -1)\over k _2}
          [\sigma _{x,b,0} + v _{T _2}(t - t _{0 _1})] -
          \xi _{2} \sigma _{x, b}(\xi _{2}) = $$
          \begin{equation}    
           - [{k _2 - 1\over k _2} ({l _{T _2}\over \sigma _{x,b,0}}
           - 1) + \xi_{2} D _{2})]\sigma _{x,b,0},
       \end{equation}
where $D _{2} = D _2(k _2, \xi _{2}) = \sigma _{x,b}/\sigma _{x,b,0}$,
$l _{T _2} = x _{T _2} - x _{T _2, 0} = \pi k_2 \psi (k _2, \xi _{2})
\sigma _{x,b,0} \leq l _f$ is the displacement of the laser beam. The
typical dependence $D _{2}$ defined by (19) is presented in Fig.4.

When $t > t _f$ then the value $\xi _{2} = \xi _{2,f} = -1$, $l _{T _2}
= l _f$, $D _{2} = \sqrt {k _2/(k _2 - 1)}$ and (19) have the maximum

       $$\Delta x _{\eta _{1-2}}| _{t > t _f}  =
       [{k _2 - 1\over k _2} + \sqrt {k _2\over k _2 -1} -$$
       \begin{equation}  
       \pi (k _2 -1) \psi (k _2, \xi _{2,f})
       ]\sigma _{x,b,0}.
       \end{equation}

The instantaneous orbit $ x _{\eta _{2}}$ will be at a distance $x
_{\eta _{2-3}} = [(k _2 -1)/k _2]\sigma _{x, \eta, 0}$ from the
motionless instantaneous orbit $x _{\eta _{3}}$ of electrons having
minimum energy and zero amplitudes of betatron oscillations when the
laser beam is stopped at the position $x _{\eta _{3}}$.

If we take into account that the instantaneous orbits of electrons
having maximum amplitudes of betatron oscillations and the minimum
energy are below the instantaneous orbits of electrons having zero
amplitudes of betatron oscillations and minimum energy, by the value
$0.28 \sigma _{x,b,0}$, at the moment of the laser beam stopping then
the total radial dispersion of the instantaneous orbits of the beam can
be presented in the form

         $$\sigma _{x,\varepsilon,f} \leq
         {k_2 - 1\over k _2} \sigma _{x,0} +
         [\sqrt {k _2\over k_2 - 1} - \pi (k _{2} -1)
         \psi (k _2, \xi _{2,f}) $$
         \begin{equation}               
         + 0.28] \sigma _{x, b,0},
         \hskip 6mm A _{T _2} > l _f, \sigma _{x,0}.
         \end{equation}

According to (21) the efficiency of the enhanced longitudinal laser
cooling is the higher the less the ratio of the spread of the initial
amplitudes of betatron oscillations to the spread of the instantaneous
orbits of the being cooled electron beam.

According to (17) and (21) the enhanced longitudinal laser cooling can
lead to a high degree of cooling of electron beams in the longitudinal
plane and a much lesser degree of heating in the transverse one. From
this it follows that the four- and six-dimensional emittances of the
beam will be decreased.

The laser beam width "{\it a}" must be higher then $\sigma _{x,
\varepsilon, f}$ in the methods of the enhanced laser cooling.

\vskip 35mm

\begin{figure}[hbt]
\includegraphics{
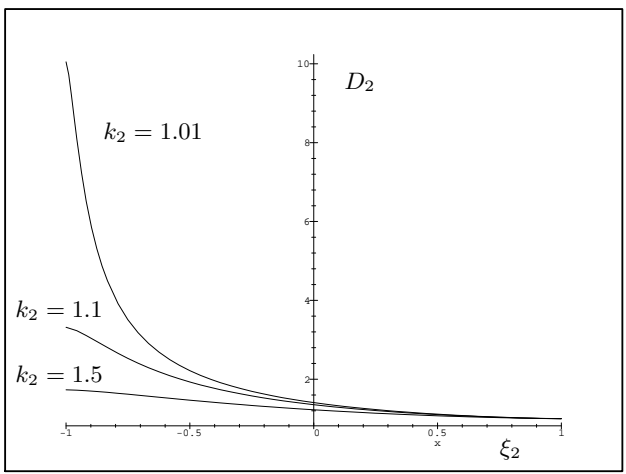}

\caption{\small \it The dependence on $\xi _{2}$ of the ratio of a
current amplitude of betatron oscillations to an initial amplitude $ D
_{2} = A /A _0 = \sigma _{x,b}/ \sigma _{x,b,0}$.}

\vskip -70mm
\hskip 10mm
{\small $D _{2}$}

\vskip 3mm
\hskip -45mm
{\small $k _2 = 1.01$}

\vskip 20mm
\hskip -70mm
{\small $k _2 = 1.1$}

\vskip 5mm
\hskip -70mm
{\small $k _2 = 1.5$}

\vskip 6mm
\hskip 50mm
{\small $\xi _{2}$}

\end{figure}
\vskip 25mm

             \subsection {Damping times}

The damping times of the electron beam in the transverse and
longitudinal methods of cooling are

        $$\tau _{x} = {2 \sigma _{x,0}\over v_{T _1}} =
        {2 \sigma _{x,0} \varepsilon \over k_{1} D _x
        \overline P},$$
        \begin{equation}     
        \tau _{s} = {2 \sigma _{\varepsilon, 0} \over
        \overline P}(1 + {\sigma _{x, b, 0}\over \sigma _{x,
        \varepsilon, 0}}),
        \end{equation}
where $D _x = (p \partial x _{\eta}/\partial p)|_{\gamma \gg
1} \simeq \varepsilon (\partial x _{\eta}/\partial \varepsilon)$ is the
local dispersion function of the storage ring at the laser beam
azimuth; $p = Mc \beta \gamma$, the momentum of the electron; $\dot x
_{\eta \,in} \simeq D _x \overline P/ \varepsilon$, $|k _1| \simeq
0.1$.

                     \section {Discussion}

The dynamics of instantaneous orbits and amplitudes of betatron
oscillations of electrons depends on the depth of deepening of their
instantaneous orbits in the laser beam and on the amplitudes. Moreover,
the being displaced laser beam begins to interact with electrons of the
beam located at different instantaneous orbits at different moments of
time and interacts for different periods of time. These features of
selective interaction of the moving laser beams lead to the enhanced
cooling of electron beams either in the transverse or longitudinal
planes.

In the method of the enhanced transverse laser cooling of electron
beams, according to (14) and (15), the degree of transverse compression
is $C _{1} = A _0/A = \sqrt{(1 + |k _1|)/ |k _1|}$ and the increase in
the spread of the instantaneous orbits of the beam (decompression), $D
_{1} \simeq C _{1} ^{2}$. At the same time in the method of the
enhanced longitudinal laser cooling, according to (17) and (21), there
is a significant decrease in the spread of instantaneous orbits of
electrons defined by the compression coefficient $C _{2,l} = \sigma
_{x, \varepsilon, 0}/ \sigma _{x, \varepsilon, f}$, and a lesser value
of increase in the amplitudes of betatron oscillations $D _{2} \simeq
\sqrt{C _{2}}$. From this it follows that cooling of electron beams
both in the transverse and longitudinal planes, in turn or
simultaneously, does not lead to their total cooling in these planes.
We can cool electron beams either in the transverse or longitudinal
planes or look for combinations of these methods of cooling with other
methods.

In the method of enhanced longitudinal laser cooling, contrary to the
transverse one, the degree of longitudinal cooling is greater than the
degree of heating in the transverse plane. That is why we can use the
emittance exchange between longitudinal and transverse planes, e.g.,
using a synchro-betatron resonance \cite {robinson, hoffman, sessler,
kihara} or dispersion coupling by additional motionless wedge-shaped
targets \cite {oneil, shoch, neufer} together with the moving target $T
_2$ and in such a way to realize the enhanced two-dimensional cooling
of the electron beam based on the longitudinal laser cooling
only\footnote{ Similar way a cooling of muon beams can be done by
material targets.}.

When the synchrotron radiation damping of electron beams in guiding
magnetic fields of lattices of storage rings is high and the radio
frequency accelerating system is switched on then we can do an
additional enhanced laser cooling of such beams in the radio frequency
buckets. Such cooling in the longitudinal plane can be produced by
using of a being displaced laser beam $T_2$ at the condition when
losses of energy by electrons in the laser beam are higher than
synchrotron radiation losses. Cooling of the electron beams in the RF
buckets is another problem to be considered elsewhere.

Notice that in  the case of the three dimensional laser cooling of
electron beams considered in \cite{zhirong} the spread of amplitudes of
betatron oscillations is small and the energy spread and the spread of
instantaneous orbits of electrons is high in another straight section
if it has a high dispersion and low $\beta$-function. That is why we
can locate the laser beam $T _2$ at this section and produce an
additional longitudinal cooling one, two or more times. Then we can use
the beam with low transverse and longitudinal emittances for generation
of spontaneous or stimulated radiation in the storage ring or extract
it for injection to a linear collider.

The enhanced transverse method of laser cooling together with the
enhanced longitudinal one (see section 2) can be used for cooling of
ion beams. Broad-band laser beam must be used in the first case and
monochromatic one in the second case. A heating of the ion beam in the
longitudinal plane in the process of the enhanced transverse cooling
will be compensated completely by its following cooling in the
longitudinal plane. If the laser beam in the transverse method of
cooling has a broad spectrum with sharp frequency edge corresponding to
an energy lesser than the minimal energy of the being cooled ion beam
then the one-dimensional method can be omitted\footnote {We can use a
"monochromatic" laser beam with a fast periodical modulation of
frequency as well.}. Ion beam will be cooled in the transverse plane
and all ions will be gathered at some energy lesser than minimal after
several cycles of transverse cooling. In this case the enhanced
transverse method of laser cooling is, at the same time, the enhanced
longitudinal method for ion beams.

Laser beams for transverse and longitudinal cooling described above can
be used for enhanced cooling of ion beams in both planes in radio
frequency buckets.

                        \section{Conclusion}

In this paper we have presented methods of enhanced laser cooling of
particle beams in transverse and longitudinal planes. These methods are
based on an universal kind of selective interaction of particles with
moving in the radial direction laser beams or media targets when the
Robinson damping criterion is not valid. We hope that these methods can
be used for cooling of ion and muon beams in storage rings intended for
the elementary particle physics.  The development and adoption of these
methods can lead to new generations of light sources of spontaneous
incoherent and stimulated radiation from optical to X-ray and
$\gamma$-ray regions based on electron and ion storage rings \cite
{pac,zhirong,esrf}. Using of circular polarized laser beams for cooling
can lead to a longitudinal polarization of stored $e^{\pm}$ beams in
storage rings with more complicated lattices \cite {bes3} - \cite
{clen}. Hard circular polarized powerful photon beams produced in the
process of the Backward Compton Scattering of laser photons by
electrons in storage rings can be used for a production in material
targets of longitudinally polarized positron beams for linear colliders
\cite {mikh, besdesy}.

We would like to thank Prof. Kwang-Je Kim and Dr. Zhirong Huang for
useful discussions.

\newpage
\addcontentsline {toc} {section} {\protect\numberline {6 \hskip 2mm
    References}}

\end{document}